\documentclass[superscriptaddress,amsfonts,amssymb,amsmath,twocolumn]{revtex4}

\usepackage{epsfig}
\usepackage{graphics}
\usepackage{graphicx}
\usepackage{amsmath}
\usepackage{color}
\usepackage{comment}
\usepackage{slashed}

\date{\today}

\newcommand{\dpar}[1]{\left(#1 \right)}
\newcommand{\dcol}[1]{\left[#1 \right]}

\begin{document}


\title{Disfavoring the Schroedinger-Newton equation}
    

\author{João V. B. da Silva}\email{jvb.silva@unesp.br}    
\affiliation{Instituto de Física Teórica, 
Universidade Estadual Paulista, Rua Dr. Bento Teobaldo Ferraz, 271, 01140-070, São Paulo, São Paulo, Brazil}

\author{Gabriel H. S. Aguiar}\email{ghs.aguiar@unesp.br}
\affiliation{Instituto de Física Teórica, 
Universidade Estadual Paulista, Rua Dr. Bento Teobaldo Ferraz, 271, 01140-070, São Paulo, São Paulo, Brazil}

\author{George E. A. Matsas}\email{george.matsas@unesp.br}
\affiliation{Instituto de Física Teórica, 
Universidade Estadual Paulista, Rua Dr. Bento Teobaldo Ferraz, 271, 01140-070, São Paulo, São Paulo, Brazil}

\pacs{} 
    

\begin{abstract}
    The main goal of this brief report is to provide some new insight into how promising the Schroedinger-Newton equation would be to explain the emergence of classicality. Based on the similarity of the Newton and Coulomb potentials, we add an electric self-interacting term to the Schroedinger-Newton equation for the hydrogen atom. Our results rule out the possibility that single electrons self-interact through their electromagnetic field.  Next, we use the hydrogen atom to get insight into the intrinsic difficulty of testing the Schroedinger-Newton equation itself and conclude that the Planck scale must be approached before sound constraints are established.  Although our results cannot be used to rule out the Schroedinger-Newton equation at all, they might be seen as disfavoring it if we underpin on the resemblance between the gravitational and electromagnetic interactions at low energies.
\end{abstract}


\maketitle


\section{Introduction}
\label{sec1} 

The emergence of the classical world from the quantum paradigm remains elusive~\cite{AHorn14}. Among the different proposals to solve this issue, let us focus on the ``quantum state reduction'' mechanism presented by Penrose~\cite{penrose96,penrose98}. In such a mechanism, gravitational self-interaction would be responsible for vanishing coherent quantum superpositions. 
One may think of it as follows. Suppose a Schroedinger-cat state where a particle is set in a coherent superposition at two different locations:  
$$
|\psi \rangle_0 
= 
\frac{1}{\sqrt 2} 
\left( 
|\, {}^\bullet \, \rangle + |\,{}_\bullet\, \rangle 
\right).
$$
Now, let us assume the {\em unorthodox} view that the two branches gravitationally interact with each other driving the system to some sort of self-entanglement:   
$$
|\psi \rangle_T 
= 
\frac{1}{\sqrt 2}
\left( 
|\, {}^\bullet \, \rangle  |\, {}_\circ \,\rangle 
+ 
|\, {}_\bullet \, \rangle  |\, {}^\circ \,\rangle
\right),
$$
with the transition time~$T$ being scaled by the gravitational interaction. Then, by tracing out over the ``immaterial'' $\{|\, {}_\circ\,\rangle, |\, {}^\circ\, \rangle \}$, one would end up with the mixed state given by the density matrix
$$
\hat{\rho}_T  
= 
\frac{1}{2} 
|\,{}^\bullet \,\rangle  \langle \, {}^\bullet \,| 
+ 
\frac{1}{2} 
|\,{}_\bullet \,\rangle  \langle \, {}_\bullet \,|.
$$
In Penrose's view,  (if confirmed) this would be a factual self-reduction responsible for the emergence of classicality. In this token, Penrose propounds that the Schroedinger equation for a particle is replaced by the so-called Schroedinger-Newton equation:
\begin{equation}\label{SNE}
    i \hbar \frac{\partial}{\partial t} \Psi(\Vec{r},t) = - \frac{\hbar^2}{2 m} \nabla^2 \Psi(\Vec{r},t) + [V(\Vec{r},t) + U_g(\Vec{r},t) ]\Psi(\Vec{r},t),
\end{equation}
where $m$ is the particle mass, $V(\Vec{r},t)$ and
\begin{equation}\label{Ug}
    U_g(\Vec{r},t) \equiv - G m^2 \int d^3\Vec{r}\,' \frac{|\Psi(\Vec{r}\,',t)|^2}{|\Vec{r} - \Vec{r}\,'|}
\end{equation}
are the external and self-interacting potentials, respectively, and $G$ is the gravitational constant. In this sense, $m |\Psi(\Vec{r},t)|^2$ plays the role of {\em real} matter density, enabling different parts of the wave function to interact with each other. Although Eq.~\eqref{SNE} is nonlinear, its results would be indistinguishable from the ones given by the Schroedinger equation for typical laboratory experiments because of the weakness of the gravitational interaction. Nevertheless, continuous efforts have been made to solve it and find its properties (see, {\em e.g.}, Refs.~\cite{diósi84,bernstein98,moroz98,AH14,bahrami14,grossardt16,bassi16,ligez23} and references therein).

On the other hand, it is well known that, at low energies, the electric and gravitational potentials only differ from each other with respect to the coupling constant, and that they have been successfully used in the Schroedinger equation to describe from the hydrogen atom~\cite{Schroedinger26} to the interference of free-falling neutrons~\cite{colella75}. Based on this, it would be ``natural'' to expect that if particles self-interact through the gravitational interaction, they should also do through the electromagnetic one~\cite{gao11}. In this token, it seems interesting to consider the Schroedinger-Newton equation for the hydrogen atom with the inclusion of the self-interacting electric potential 
\begin{equation}\label{Ue}
    U_e(\Vec{r},t) \equiv k_e e^2 \int d^3\Vec{r}\,' \frac{|\Psi(\Vec{r}\,',t)|^2}{|\Vec{r} - \Vec{r}\,'|},
\end{equation}
where $k_e$ is the Coulomb constant and $e$ is the fundamental charge.

This paper is organized as follows. In Sec.~\ref{sec2}, we introduce the Schroedinger-Newton equation for the hydrogen atom modified by the addition of the electric self-interacting potential. In Sec.~\ref{sec3}, we solve it numerically and compare the outputs with experimental data. In  Sec.~\ref{sec4}, we use the hydrogen atom to drive some conclusions about the intrinsic difficulty of testing the Schroedinger-Newton equation itself. In Sec.~\ref{sec5}, we discuss the results and present our conclusions.


\section{The hydrogen atom under self-interaction}
\label{sec2}

The Schroedinger equation for the hydrogen atom is
\begin{equation}\label{HAE}
    i \hbar \frac{\partial}{\partial t} \Psi(\Vec{r},t) = - \frac{\hbar^2}{2 \mu} \nabla^2 \Psi(\Vec{r},t) + V_T(r) \Psi(\Vec{r},t),
\end{equation}
where $\mu \equiv m_e m_p/(m_e + m_p)$, $m_e$ and $m_p$ are the electron and proton masses, respectively, and
\begin{equation}
V_T(r)\equiv V_g (r) + V_e (r)
    \label{VT}
\end{equation}
with $V_g (r) = - G m_e m_p/r$ and $V_e(r) = -{k_e e^2}/{r}$. (We recall that the corresponding spectroscopy is completely determined by the spherically symmetric eigenfunctions.)

Our main goal is to compare the results given by Eq.~\eqref{HAE} with the ones provided by the Schroedinger-Newton equation modified by the addition of the electric self-interacting potential:
\begin{equation}\label{MHAE}
    i \hbar \frac{\partial}{\partial t} \Psi(\Vec{r},t) = - \frac{\hbar^2}{2 \mu} \nabla^2 \Psi(\Vec{r},t) + \dcol{V_T(r) + U_T(\Vec{r},t)} \Psi(\Vec{r},t),
\end{equation}
where 
\begin{equation}
U_T(\Vec{r},t) \equiv U_g(\Vec{r},t) + U_e(\Vec{r},t).
    \label{UT}
\end{equation}
Clearly, the electric terms supersede the gravitational ones by many orders of magnitude and, thus, the latter could be omitted at this point. Still, we opted to maintain them for the sake of our further discussion. We shall note that Eq.~\eqref{MHAE} with $V_g (r) = U_g(\vec r, t)=0$ was investigated in the past in Ref.~\cite{afanasiev94} but finding no bound states for the hydrogen atom. This is at odds with our results and with the ones obtained in Refs.~\cite{rañada81,biguaa20}, which considered the Dirac equation with electric self-interaction to resolve the hydrogen atom. Interestingly enough, by comparing the results of Refs.~\cite{rañada81,biguaa20} with ours, we see that the self-interacting term plays a more significant role in the nonrelativistic realm than in the relativistic one. In particular, the electron would be five times less bound in the former than in the latter case. (This may explain why the authors of Ref.~\cite{afanasiev94} were unable to find bound states for the hydrogen atom.) Most importantly, however, is the fact that Refs.~\cite{rañada81,afanasiev94,biguaa20} consider electric self-interaction in the hydrogen atom under a quite different mindset with respect to ours; quoting Biguaa and Kassandrov: 
\begin{quote}  
(A) ``From the point of view of quantum ideas, the electron in the hydrogen atom represents a spatially distributed system, according to the probability density $\Psi^{\dagger} \Psi$, whose elements interact both with the field of the nucleus and between each other." 
\end{quote}
In order to fit their results with experimental data, they appeal to a sort of finite ``classical renormalization" procedure. As expected, this is not enough to recover consistency with experiments. We shall emphasize that standard quantum mechanics does not assert (A). Instead, it states that $|\Psi(\Vec{r},t)|^2$ is the probability density of finding the electron {\em when} its position is measured by some apparatus. Also, allegations that radioactive corrections of quantum field theory would endorse self-interaction of single-particle states are unfounded. 

Here we compare the results delivered by Eq.~\eqref{MHAE} with the ones provided by quantum mechanics taken at its face value. Let us look for stationary spherically symmetric solutions 
\begin{equation*}
    \Psi(\Vec{r},t) = \psi(r) e^{- i E t/\hbar}
\end{equation*}
for Eq.~\eqref{MHAE}. In this case, Eq.~\eqref{MHAE} can be cast as
\begin{equation}\label{MHAES}
    - \frac{\hbar^2}{2 \mu} \frac{1}{r} \frac{d^2}{d r^2} \dcol{r \psi(r)} + \dcol{V_T(r) + W_T(r)} \psi(r) = E \psi(r),
\end{equation}
where
\begin{equation}\label{W}
    W_T(r) = 4 \pi  \alpha \dpar{\frac{1}{r} \int\limits_0^r dr' r'^2 |\psi(r')|^2 + \int\limits_r^\infty dr' r' |\psi(r')|^2}
\end{equation}
and $\alpha \equiv -Gm_e m_p + k_e e^2$.
In order to write it, we have used (see, {\em e.g.,} Ref.~\cite{jackson99})
\begin{equation*}
    \frac{1}{|\Vec{r} - \Vec{r}\,'|} = 4 \pi \sum\limits_{l = 0}^\infty \sum\limits_{m = -l}^l \frac{1}{2l + 1} \frac{r_<^l}{r_>^{l + 1}} Y_{lm}^*(\theta',\phi') Y_{lm}(\theta,\phi),
\end{equation*}
where $r_< \equiv \min(r,r')$, $r_> \equiv \max(r,r')$, and  $Y_{lm}(\theta,\phi)$ are the spherical harmonics. Equation~\eqref{MHAES} can be simplified by defining $\phi(r) \equiv r \psi(r)$, in which case it becomes
\begin{equation}\label{MHAESP}
    - \frac{\hbar^2}{2 \mu} \frac{d^2}{d r^2} \phi(r) + \dcol{V_T(r) + W_T(r)} \phi(r) = E \phi(r),
\end{equation}
where
\begin{equation}\label{WP}
    W_T(r) = 4 \pi \alpha \dpar{\frac{1}{r} \int\limits_0^r dr' |\phi(r')|^2 + \int\limits_r^\infty dr' \frac{|\phi(r')|^2}{r'}}
\end{equation}
and we recall that the probability of finding the electron in a spherical shell with inner and outer radius $r_i$ and $r_o$, respectively, is
\begin{equation}\label{F}
    P(r_i,r_o) = \int\limits_{r_i}^{r_o} dr F(r)
    \quad
    \text{with}
    \quad
    F(r) \equiv 4 \pi |\phi(r)|^2.
\end{equation}

Although the quest for stationary spherically symmetric solutions simplifies the problem, Eq.~\eqref{MHAESP} is still nontrivial. In order to deal with it, we shall solve it numerically by adapting the code of Ref.~\cite{bernstein98}.


\section{Numerical procedure and results}
\label{sec3}

The code employed to numerically solve Eq.~\eqref{MHAESP} is included in the Supplemental Material. This is a user-friendly adaptation of the one developed in Ref.~\cite{bernstein98}, where the external potential $V_T(r)$ was included and $U_g(\Vec{r},t)$ was replaced by $U_T(\Vec{r},t)$. (Furthermore, the convergence procedure was somewhat improved.) Briefly speaking, we depart from a test function, use it to calculate $W_T(r)$ in Eq.~\eqref{WP}, and replace the result in Eq.~\eqref{MHAESP} to numerically evaluate the eigenfunctions~$\phi(r)$ (and corresponding eigenvalues~$E$). Then, depending on the eigenvalue~$E$ we are interested in, we select the associated eigenfunction~$\phi(r)$, using it in the next round as a new test function to recalculate $W_T(r)$ and so on. The process is repeated until the eigenenergy~$E$ of interest converges.

As a consistency check, we verify at the end that the obtained eigenfunctions and corresponding eigenenergies do satisfy Eq.~\eqref{MHAESP}. We have also confirmed that the code reproduces in good approximation the experimental results for the energy levels of the hydrogen atom in the absence of self-interaction~\cite{tiesinga21} -- see Table~\ref{table1}. (By ``Level~$n$'', $n=1,2,3,\ldots$, we  implicitly mean level $n\, s^{1/2}$, since the solutions are spherically symmetric.)
\begin{table}[htbp]
    \centering
    \begin{tabular}{| c | c | c |} 
     \hline
     \, Level \, & \, Numerical (eV) \, & \, Experimental (eV) \, \\ [0.5ex]
     \hline\hline
     $1$ & $-13.593$ & $-13.598$\\ 
     $2$ & $-3.3993$ & $-3.3996$\\
     $3$ & $-1.5109$ & $-1.5109$\\ [1ex] 
     \hline
    \end{tabular}
    \caption{Comparison of the output of the three smallest eigenenergies given by Eq.~\eqref{HAE} against the experimental data (restricted to $5$ significant figures). }
    \label{table1}
\end{table}

Now, we are in a position to compare the impact of considering self-interaction on the spectroscopy of the hydrogen atom as ruled by the modified Schroedinger-Newton equation~\eqref{MHAESP}. We shall stress here that any deviations from the experimental data must be imputed to the electric self-interaction due to the huge dominance of $k_e e^2$ over $Gm_e m_p$. In Table~\ref{table2}, we exhibit the three smallest eigenenergies for the hydrogen atom. It becomes clear, in particular, that the ionization energy is at odds with the experimental result due to the electric self-interaction. Indeed, the whole hydrogen atom spectroscopy is ruined, as can be seen in Table~\ref{table3}. 

\begin{table}[htbp]
    \centering
    \begin{tabular}{| c | c | c |} 
     \hline
     \, Level \, & \, Numerical (eV) \, & \, Experimental (eV) \, \\ [0.5ex]
     \hline\hline
     $1$ & $-1.2561$ & $-13.598$ \\ 
     $2$ & $-0.21601$ & $-3.3996$ \\
     $3$ & $-0.074618$ & $-1.5109$ \\ [1ex] 
     \hline
    \end{tabular}
    \caption{Comparison of the three smallest eigenenergies, numerically obtained through Eq.~\eqref{MHAESP}, against experimental data (restricted to $5$ significant figures).}
    \label{table2}
\end{table}

\begin{table}[htbp]
    \centering
    \begin{tabular}{| c | c | c |} 
     \hline
     \, Transition \, & \, Numerical (eV) \, & \, Experimental (eV) \, \\ [0.5ex] 
     \hline\hline
     $3 \to 2$ & $-0.14139$ & $-1.8887$ \\
     $2 \to 1$ & $-1.0400$ & $-10.199$ \\
     $3 \to 1$ & $-1.1814$ & $-12.087$ \\ [1ex] 
     \hline
    \end{tabular}
    \caption{Comparison of the energy gap for three transitions, numerically obtained through Eq.~\eqref{MHAESP}, against experimental data (restricted to $5$ significant figures). By ``$n_i \to n_f$," we mean the electronic transition from the $n_i$-th to the $n_f$-th energy eigenstate. }
    \label{table3}
\end{table}

Let us note that the energy levels assuming electric self-interaction shown in Table~\ref{table2} do not follow the usual Bohr relationship $E_n = E_1/n^2$. This is expected since the self-interacting term should screen differently the proton charge depending on the quantum state.

\begin{figure}[htbp]
    \includegraphics[width=75mm]{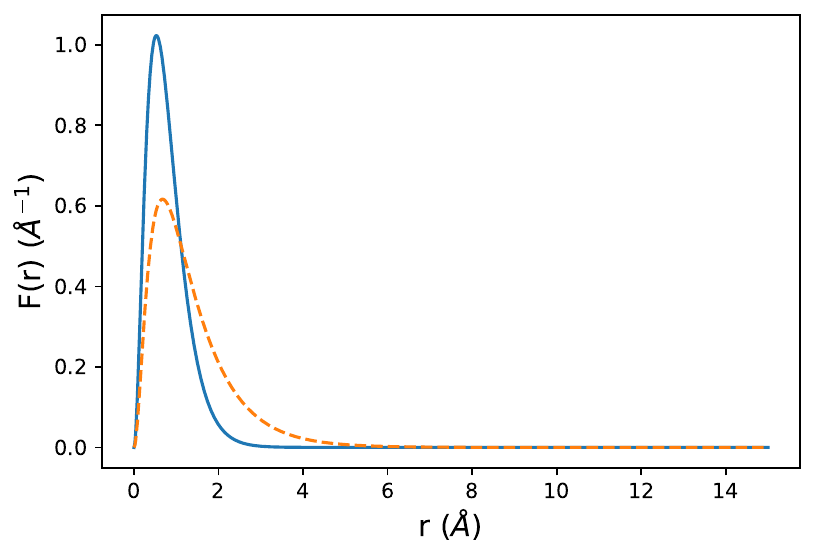}
    \caption{Plot of the radial probability density $F(r)$ for the ground state obtained from Eq.~\eqref{HAE} (solid line) and Eq.~\eqref{MHAESP} (dashed line).}
    \label{fig1}
\end{figure}

\begin{figure}[htbp]
    \includegraphics[width=75mm]{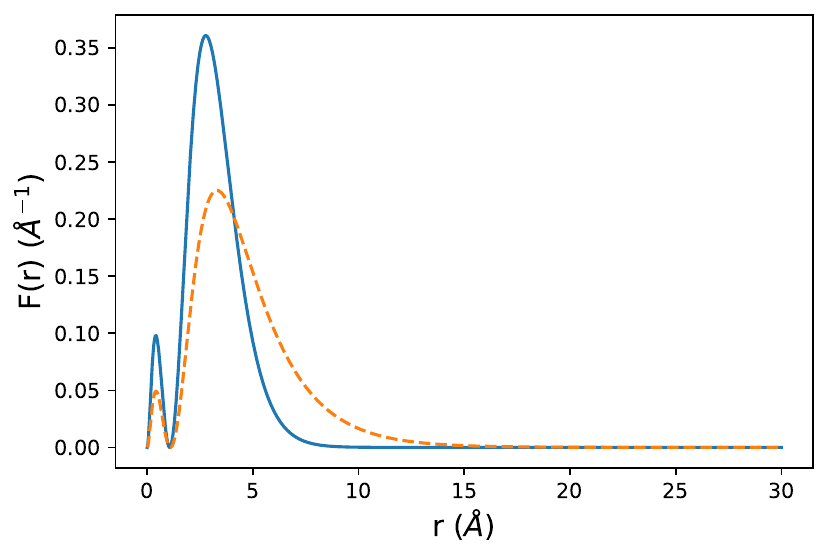}
    \caption{Plot of the radial probability density $F(r)$ for the first excited state obtained from Eq.~\eqref{HAE} (solid line) and Eq.~\eqref{MHAESP} (dashed line).}
    \label{fig2}
\end{figure}

\begin{figure}[h!]
    \includegraphics[width=75mm]{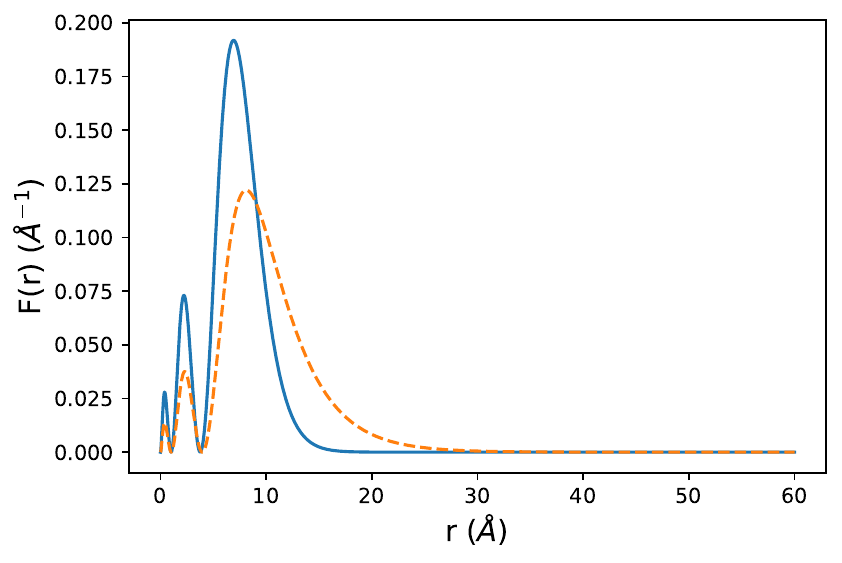}
    \caption{Plot of the radial probability density $F(r)$ for the second excited state obtained from Eq.~\eqref{HAE} (solid line) and Eq.~\eqref{MHAESP} (dashed line).}
    \label{fig3}
\end{figure}

In Figs.~\ref{fig1}, \ref{fig2}, and \ref{fig3}, we compare the radial probability densities $F(r)$ [see Eq.~\eqref{F}] associated with the eigenfunctions of the three smallest eigenenergies given by the Schroedinger equation~\eqref{HAE} and modified Schroedinger equation~\eqref{MHAESP}. We see that the electric self-interaction pushes the probability density farther from the nucleus, in agreement with the larger eigenenergies found for the corresponding eigenfunctions displayed in Table~\ref{table2}.

Although the results above experimentally discard the existence of electromagnetic self-interaction and may be understood as disfavoring the Schroedinger-Newton equation (based on the similarity of the Newton and Coulomb potentials), they cannot discard gravitational self-interaction at all. In order to test it itself, we must analyze situations where the gravitational self-interaction turns out to be dominant. In the next section, we use the hydrogen atom to gain insight into how challenging this can be. 


\section{The hydrogen atom restricted to gravitational self-interaction}
\label{sec4}

Let us start replacing Eq.~\eqref{MHAE} by 
\begin{equation}\label{MHAE2}
    i \hbar \frac{\partial}{\partial t} \Psi(\Vec{r},t) = - \frac{\hbar^2}{2 \mu} \nabla^2 \Psi(\Vec{r},t) + \dcol{V_T(r) + U_g(\Vec{r},t)} \Psi(\Vec{r},t),
\end{equation}
where the electric self-interaction was promptly discarded in the face of the previous results. It is fair to expect that the presence of $U_g(\Vec{r},t)$ should effectively increase the nucleus mass by no more than a factor of the electron mass. This comes from Sec.~\ref{sec3} where $U_e(\Vec{r},t)$ was shown to effectively screen part of the proton charge. Note that by driving $Z \to 0.3 \; Z$, the ionization energy
\begin{equation}\label{IonizationEnergy}
E_1 =- \frac{Z^2 \mu}{2\hbar^2 n^2} (k_e e^2)^2
\quad
(Z=1,\; n=1)
\end{equation}
would change about $ -13.598~{\rm eV} \to -1.2561~{\rm eV} $ 
(see Table~\ref{table2}). 
In this token, the ionization energy, as ruled by Eq.~\eqref{MHAE2}, can be estimated to be of the order
\begin{equation}\label{IonizationEnergyPrime}
E'_1  \sim -\frac{Z^2 \mu}{2\hbar^2 n^2} (k_e e^2 + G m_p m_e) ^2 
\quad
(Z=1,\; n=1),
\end{equation}
where any enhancement of the proton mass due to the gravitational self-interaction of the electron will be neglected for now since it would only lead to a tiny correction to the $Gm_e m_p$ term. Hence, gravity would affect the ionization energy by 
\begin{equation} \label{ratio}
\frac{{E'}_1 - {E}_1}{E_1} \sim 2 \left( \frac{G m_p m_e}{k_e e^2}  \right),
\end{equation}
corresponding to $1$ part in $10^{39}$ parts. To put this in context, we recall that the value recommended by CODATA for the Newtonian constant has only six significant digits: $G = 6.67430(15) \times 10^{-11} \ \text{m}^{-3} \cdot \text{kg}^{-1} \cdot \text{s}^{-2}$~\cite{tiesinga21}. We note that in order to drive the right-hand side of Eq.~\eqref{ratio} to approach the unity, we should have had $m_e m_p \sim (M_P/10)^2$, where 
$$
M_P\equiv{\sqrt{\hbar c/G\,}\approx 2\times 10^{-5}} {\rm g}
$$ 
is the Planck mass. 

Eventually, the paramount obstacle faced by every tentative to test gravitational self-interaction is that it must approach the quantum gravity regime by combining coherently a large number of particles to act as a single quantum particle with mass $m \lesssim M_P$. It is unclear, however, in what sense Schroedinger and other quantum wave equations would be approximately valid close to the Planck scale. Relativistic, 
\begin{equation}\label{WaveEquationsR}
\Box \phi + \lambdabar^{-2} \phi=0,
\quad
i \;\slashed{\partial} \,\psi - \lambdabar^{-1} \psi=0,
\end{equation}
and nonrelativistic, 
\begin{equation}\label{WaveEquationsNR}
i {\partial_t \phi} + (c \lambdabar/2) \nabla^2 \phi =0,
\end{equation}
wave equations are solely characterized by the (reduced) Compton wavelength $\lambdabar = \hbar/(mc)$. The closer $m$ approaches~$M_P$, the closer $\lambdabar$ approaches the Planck length: 
$$
\lambdabar \sim L_P\equiv{\sqrt{\hbar G/c^3\,}}.
$$ 
Hence, in order to write Eqs.~\eqref{WaveEquationsR}-\eqref{WaveEquationsNR} for Planck-mass particles, one must presuppose that classical spacetimes are reliable at such distance scales, which does not seem to be the case. 

In relativistic spacetimes, spatial distances can be solely measured using bona-fide classical clocks. By {\em bona-fide classical clocks} we mean ``pointike'' apparatuses that ascribe the same real number (time interval) to any given arbitrarily-close-causally-connected pair of events they visit regardless of the state of motion and past history (as defined by relativity). Thus, in order to measure $\lambdabar \sim L_P$, there must exist bona-fide clocks with size $L_\text{clock} < L_P$ and accuracy  
$$
\delta t \lesssim T_P \equiv {\sqrt{\hbar G/c^5\,}},
$$ 
in which case one would end with a ruler with accuracy $\delta l \equiv c \delta t < \lambdabar$. On the other hand, quantum mechanics teaches us that clocks with such an accuracy would involve Planck-scale energies: $E_\text{clock} \gtrsim \hbar/ \delta t$~\cite{Schumacher10}, leading to a clock mass
$$
M_\text{clock} \gtrsim M_P.
$$  
It happens, however, that according to Thorne's hoop conjecture such clocks should collapse into black holes since they would be small enough to fit in spheres with the corresponding Schwarzschild radii~$R^S_\text{clock}$ (see, {\it e.g.,} Ref.~\cite{Misner73}): 
$$
L_\text{clock} < L_P < 2G M_P/c^2 \lesssim 2G M_\text{clock}/c^2 \equiv R^S_\text{clock}. 
$$ 
Not only black holes are not supposed to function as clocks but such apparatuses would disturb the background spacetime raising doubts on the validity of Eq.~\eqref{SNE} written for the Galilean spacetime.

As a result, any proposal to assess the Schroedinger-Newton equation must approach the Planck scale but still not cross the line where bona-fide clocks are absent and Galilean spacetime cannot be assumed. In Ref.~\cite{grossardt16}, it is suggested the employment of Paul (ionic) traps to confine osmium disks of $10^{-9}~\text{g}$ subjected to a harmonic potential. Traces of gravitational self-interaction would be fingerprinted in the corresponding energy spectrum. For this purpose, it would be necessary to reach temperatures of a few milikelvin. This has already been achieved in Paul traps for masses of $10^{-14}~\text{g}$ \cite{bykov19,dania21,penny23} but not for ones of $10^{-9}~\text{g}$ yet. (We address to Ref.~\cite{ballestero22} for an actual review about particle trapping technology.)


\section{Conclusions}
\label{sec5}

The Schroedinger-Newton equation was proposed as a trial to recover the classical world from the quantum paradigm. Briefly speaking, particles would decohere due to gravitational self-interaction. On the other hand, at low energies, the gravitational and electric potentials are formally the same (only differing due to the universality of the coupling constant in the gravitational case). This has motivated us to revisit the hydrogen atom, assuming that the electron self-interacts through its proper charge. We quested for spherically symmetric stationary solutions and compared the outputs with the experimental data. We have obtained that, according to the modified Schroedinger equation~\eqref{MHAESP}, the ionization energy of the hydrogen atom and the corresponding spectroscopy data are at odds with each other, ruling out the possibility that electrons self-interact through their charge. Although our results only refer to the electromagnetic interaction, they may be seen as disfavoring the Schroedinger-Newton equation in the sense that if a single electron would ``perceive" itself through its mass distribution, it would be ``natural" to argue that the same would be true concerning its charge distribution. In the end, we assessed some intrinsic difficulties of testing the Schroedinger-Newton equation as such.

%
%


\acknowledgments

The authors would like to acknowledge discussions with Caslav Brukner, Anne-Catherine de la Hamette, and Viktoria Kabel. J.~S. and G.~A. were fully supported by the Coordination for the Improvement of Higher Education Personnel (CAPES) and S\~ao Paulo Research Foundation (FAPESP) under grants 88887.637540/2021-00 and~2022/08424-3, respectively. G.~M. was partially supported by the National Council for Scientific and Technological Development and FAPESP under grants 301508/2022-4 and 2022/10561-9, respectively.



\begin{thebibliography}{}

    \bibitem{AHorn14}
    M. Arndt and K. Hornberger, Testing the limits of quantum mechanical superpositions, Nature Phys. \textbf{10}, 271 (2014).

    \bibitem{penrose96}
    R. Penrose, On gravity’s role in quantum state reduction, Gen. Relativ. Gravit. \textbf{28}, 581 (1996).

    \bibitem{penrose98}
    R. Penrose, Quantum computation, entanglement and state reduction, Phil. Trans. R. Soc. A \textbf{356}, 1927 (1998).

    \bibitem{diósi84}
    L. Diósi, Gravitation and quantum-mechanical localization of macro-objects, Phys. Lett. A \textbf{105}, 199 (1984).

    \bibitem{bernstein98}
    D. H. Bernstein, E. Giladi, and K. R. W. Jones, Eigenstates of the gravitational Schroedinger equation, Mod. Phys. Lett. A \textbf{13}, 2327 (1998).

    \bibitem{moroz98}
    I. M. Moroz, R. Penrose, and P. Tod, Spherically-symmetric solutions of the Schroedinger-Newton equations, Class. Quantum Grav. \textbf{15}, 2733 (1998).

    \bibitem{AH14}
   C. Anastopoulos and B. L. Hu, Problems with the Newton–Schroedinger equations, New J. Phys. \textbf{16}, 085007 (2014).

    \bibitem{bahrami14}
    M. Bahrami, A. Grossardt, S. Donadi, and A. Bassi, The Schroedinger–Newton equation and its foundations, New J. Phys. \textbf{16}, 115007 (2014).

    \bibitem{grossardt16}
    A. Grossardt, J. Bateman, H. Ulbricht, and A. Bassi, Optomechanical test of the Schroedinger-Newton equation, Phys. Rev. D \textbf{93}, 096003 (2016).

    \bibitem{bassi16}
    A. Grossardt, J. Bateman, H. Ulbricht, and A. Bassi, Effects of Newtonian gravitational self-interaction in harmonically trapped quantum systems, Sci. Rep., \textbf{6}, 30840 (2016).

    \bibitem{ligez23}
    R. Ligez, R. B. MacKenzie, V. Massart, M. B. Paranjape, and U. A. Yajnik, What is the gravitational field of a mass in a spatially nonlocal quantum superposition? Phys. Rev. Lett. \textbf{130}, 101502 (2023).

    \bibitem{Schroedinger26}
    E. Schroedinger, Quantisierung als Eigenwertproblem, Ann. Phys. \textbf{384}, 361 (1926).

    \bibitem{colella75}
    R. Colella, A. W. Overhauser, and S. A. Werner, Observation of gravitationally induced quantum interference, Phys. Rev. Lett. \textbf{34}, 1472 (1975).

    \bibitem{gao11}
    S. Gao, The wave function and quantum reality, AIP Conf. Proc. \textbf{1327}, 334 (2011).

    \bibitem{afanasiev94}
    G. N. Afanasiev, V. G. Kartavenko, and A. B. Pestov, Coulomb self-action effect on shift of atomic levels, Bull. Rus. Acad. Sci. Phys. \textbf{58}, 738 (1994).
    
    \bibitem{rañada81}
    A. F. Rañada and J. M. Uson, Bound states of a classical charged nonlinear Dirac field in a Coulomb potential, J. Math. Phys. \textbf{22}, 2533 (1981).

    \bibitem{biguaa20}
    L. V. Biguaa and V. V. Kassandrov, The hydrogen atom: consideration of the electron self-field, Phys. Part. Nuclei \textbf{51}, 965 (2020).

    \bibitem{jackson99}
    J. D. Jackson, {\em Classical Electrodynamics}, 3rd ed. (John Wiley, New York, 1999).

    \bibitem{tiesinga21}
    E. Tiesinga, P. J. Mohr, D. B. Newell, and B. N. Taylor, CODATA recommended values of the fundamental physical constants: 2018, Rev. Mod. Phys. \textbf{93}, 025010 (2021).

     \bibitem{Schumacher10}
   B. Schumacher and M. Westmoreland, {\it Quantum Processes Systems, and Information} (Cambridge University Press, Cambridge, 2010).

   \bibitem{Misner73}
   C. W. Misner, K. S. Thorne, and J. A. Wheeler, {\it Gravitation} (Freeman, San Francisco, 1973).

    \bibitem{bykov19}
    D. S. Bykov, L. Dania, P. Mestres, and T. E. Northup,  Laser cooling of secular motion of a nanoparticle levitated in a Paul trap for ion-assisted optomechanics in {\em Proc. SPIE 11083, Optical Trapping and Optical Micromanipulation XVI, 110831D} (2019).

    \bibitem{dania21}
    L. Dania, D. S. Bykov, M. Knoll, P. Mestres, and T. E. Northup, Optical and electrical feedback cooling of a silica nanoparticle levitated in a Paul trap, Phys. Rev. Res. \textbf{3}, 013018 (2021).

    \bibitem{penny23}
    T. W. Penny, A. Pontin, and P. F. Barker, Sympathetic cooling and squeezing of two colevitated nanoparticles, Phys. Rev. Res. \textbf{5}, 013070 (2023).

    \bibitem{ballestero22}
    C. Gonzalez-Ballestero, M. Aspelmeyer, L. Novotny, R. Quidant, and O. Romero-Isart, Levitodynamics: Levitation and control of microscopic objects in vacuum, Science, \textbf{374}, 168 (2021).

\end{thebibliography}
\end{document}